\documentclass[12pt,preprint]{emulateapj}
\usepackage{amsmath, natbib, placeins}
\begin{document}


\title{Circular Polarization of Pulsar Wind Nebulae and the Cosmic-Ray Positron Excess}
\author{Tim Linden$^{1}$}
\affil{$^1$ The Kavli Institute for Cosmological Physics, University of Chicago, Chicago, IL 60637, USA}
\shortauthors{}

\begin{abstract}
Recent observations by the PAMELA and AMS-02 telescopes have uncovered an anomalous rise in the positron fraction at energies above 10~GeV. One possible explanation for this excess is the production of primary electron/positron pairs through electromagnetic cascades in pulsar magnetospheres. This process results in a high multiplicity of electron/positron pairs within the wind-termination shock of pulsar wind nebula (PWN). A consequence of this scenario is that no circular polarization should be observed within PWN, since the contributions from electrons and positrons exactly cancel. Here we note that current radio instruments are capable of setting meaningful limits on the circular polarization of synchrotron radiation in PWNs, which observationally test the model for pulsar production of the local positron excess. The observation of a PWN with detectable circular polarization would cast strong doubt on pulsar interpretations of the positron excess, while observations setting strong limits on the circular polarization of PWN would lend credence to these models. Finally, we indicate which pulsar wind nebulae are likely to provide the best targets for observational tests of the AMS-02 excess.
\end{abstract}

\section{Introduction}
\label{sec:introduction}

In 2009, the PAMELA collaboration reported an anomalous rise in the ratio of cosmic-ray positrons to cosmic-ray electrons at energies above $\sim$~10~GeV~\citep{2009Natur.458..607A}. This observation sparked a firestorm of novel theoretical models, as the conventional framework of cosmic-ray propagation indicates that positrons are produced as cosmic-ray secondaries from hadronic interactions in the galaxy - and thus the antimatter component of cosmic-ray leptons should fall smoothly with energy. This observation has been confirmed by both additional PAMELA data~\citep{2013JETPL..96..621A} and recent results from the AMS-02 experiment, which greatly improved the statistical errors on the positron measurement~\citep{2013PhRvL.110n1102A}. AMS-02 observations also extended the measurement to higher energies, and observed a slight softening in the rising positron spectrum.

Several mechanisms have been posited to explain the rising positron fraction. The majority of models employ dark matter annihilation in order to produce a primary flux of e$^+$e$^-$ pairs. These models generically require massive, leptophilic dark matter candidates with a cross-section significantly exceeding the thermal cross-section \citep[see e.g.][]{2009PhRvD..79a5014A, 2009PhRvD..80l3518C}. AMS-02 observations indicating a softening in the positron fraction above 200~GeV have made these models more difficult to fit to observations, as most leptophilic dark matter models predict the slope of the positron fraction to steepen until a cutoff is found at the mass of the dark matter particle~\citep{2013PhRvD..88b3013C}. Additionally, strong limits from $\gamma$-ray observations of dwarf spheroidal galaxies and the galactic center have ruled out many dark matter models of the positron excess~\citep{2011PhRvL.107x1302A, 2013APh....46...55H}.

Alternatively, a population of young pulsars may accelerate leptons to high energies and produce the primary positron flux necessary to explain the rising positron fraction. Calculations show that these pulsars can provide the entirety of the observed e$^+$e$^-$ flux, even if only a small percentage of their spin-down luminosity is converted into the acceleration of e$^+$e$^-$ in the pulsar magnetosphere~\citep{2009JCAP...01..025H, 2012CEJPh..10....1P}. The positron flux observed at the solar position may be due to the entire galactic population of young pulsars~\citep{2009PhLB..678..283B}, or may be dominated by a few, nearby sources~\citep{2012CEJPh..10....1P}. In the latter case, the contributions of nearby pulsars to the total positron flux may be observable as an anisotropy in the total cosmic-ray lepton flux~\citep{2013ApJ...772...18L}.

Although the fact that pulsars accelerate copious high energy electrons makes them a convincing explanation of the positron excess, significant uncertainties remain in pulsar emission modeling which may substantiate or disfavor the pulsar explanation. These uncertainties can be broken down into two major questions. First, is the multiplicity of e$^+$e$^-$ pairs produced by a single seed electron sufficient to wash out the initial matter-antimatter asymmetry? Second, are the e$^+$e$^-$ pairs produced by the candidate pulsar able to effectively escape from the confining magnetic fields of the PWN before losing their energy to synchrotron and inverse-Compton scattering~\citep{2011heep.conf..624B}?

In this \emph{letter} we show that measurements of the circular polarization in PWN can test both of these uncertainties. The high linear polarization observed in multiple PWN indicate that detectable levels of circular polarization should be observed in PWN sources. However, circular polarization has yet to be detected in any PWN. One solution to this discrepancy is that positrons and electrons have equal energy densities within PWN, canceling their relative contributions to the observed circular polarization. We show that low-frequency observations by the LOFAR telescope may be capable of placing strong constraints on the positron fraction within PWN. These observations would be capable of either bringing credence to, or ruling out, models where pulsar emission explains the rising positron fraction observed by PAMELA and AMS-02.

\section{Pulsar Emission Models}
\label{sec:pulsars}
Pulsars have been studied extensively at radio energies over the last five decades. In addition to producing bright, pulsating emission near the surface of the neutron star, pulsars also produce relativistic winds of high-energy particles which carry away most of their spin-down energy. These high-energy particles power a bright synchrotron nebulae surrounding the nascent pulsar, and are confined for some time after the supernovae by the interaction of the PWN shockwave with the reverse shock of the supernovae remnant~\citep{2006ARA&A..44...17G}. Observations have detected significant linear and circular polarization from the observed radio point source~\citep{1998MNRAS.300..373H}.

Models of pulsar emission generically agree that the synchrotron emission from near the pulsar surface is generated by electrons which are ``boiled" off of the neutron star surface by intense electric and magnetic fields, and then accelerated to high-energies~\citep{1971ApJ...164..529S, 1982ApJ...252..337D}. When a threshold energy is reached these electrons produce $\gamma$-rays via synchrotron radiation which in turn produce e$^+$e$^-$ pairs~\citep{1971ApJ...164..529S, 1982ApJ...252..337D}. This process can potentially repeat many times, producing a large number of lepton pairs from a single seed electron. The number of e$^+$e$^-$ pairs produced by the average seed electron is termed the multiplicity of the pulsar, and is described by the variable $\eta$. If $\eta$ is very large, then the final state contains an almost equivalent number of e$^+$e$^-$, and the initial matter-antimatter asymmetry is washed out.

However, models differ on the region in which e$^+$e$^-$ pairs are produced: polar cap (PC) models predict a lepton pair-production region within tens of km of the NS surface, slot gap (SG) models predict a lepton pair-production region which extends in a thin slot above the PC and accelerates electrons up until the last open field line~\citep{1979ApJ...231..854A}, outer gap (OG) models, predict lepton acceleration which begins farther away from the pulsar when the magnetic field becomes perpendicular to the pulsar rotation axis.

Among the many differences between these models are divergent predictions for the e$^+$e$^-$ multiplicity. Estimates range from an expected multiplicity, $\eta$, of 1-10 in the case of polar-cap models~\citep{low_multiplicities_in_polar_cap_models}, to multiplicities on the order of 10$^5$ in the case of outer gap models~\citep{2010ApJ...715.1318T}. In cases where the multiplicity is extremely high, and copious e$^+$e$^-$ pairs escape into the PWN, the matter-antimatter asymmetry is essentially washed out~\citep{2007ApJ...658.1177D}. It is these leptons which power the bright synchrotron emission inside the PWN (see \citet{2006ARA&A..44...17G} and \citet{2009ASSL..357..421K} for complete reviews). Recent models of pulsars observed by the Fermi-LAT indicate that high-latitude models such as the outer gap or slot gap are most consistent with $\gamma$-ray observations~\citep{2014arXiv1403.3849P}.

\section{Propagation of Leptons in PWN}
\label{sec:leptonpropagation}

A second uncertainty involves the mechanism by which energetic leptons escape from the confining magnetic fields of the PWN and enter the interstellar medium (ISM). While copious e$^+$e$^-$ pairs may be produced in pulsar magnetospheres, they quickly lose energy to the very strong magnetic fields in the pulsar magnetosphere, and are unable to provide the TeV e$^+$e$^-$ necessary to explain the $\gamma$-ray signals observed from PWN. Instead, observed PWN emission comes from the subsequent interaction (and reacceleration) of these e$^+$e$^-$ pairs at the site where the relativistic e$^+$e$^-$ wind collides with the slowly expanding supernova ejecta~\citep{2011heep.conf..624B}. Once these leptons are accelerated to high energies, they may explain the rising positron fraction observed by PAMELA and AMS-02. 

However, there are several difficulties with this scenario. First, $\gamma$-ray observations indicate that the lepton population is best fit by a broken power-law that breaks at approximately 50~GeV, and is uncharacteristic of a simple shock-acceleration model~\citep{2010arXiv1005.1831B}. This observation has also been seen in multi wavelength observations of the Vela pulsar, where the X-Ray and TeV observations have been difficult to reconcile with GeV and radio observations, indicating the existence of different emission models on different energy scales \citep{2008ApJ...689L.125D}. Several models have been proposed explain this inconsistency including: the acceleration of an additional population of thermal electrons at the termination shock~\citep{1996MNRAS.278..525A}, a second acceleration mechanism which can produce a low-energy lepton population~\citep{2002ASPC..271...99G}, or are indicative of alterations in the theory of acceleration at termination shocks~\citep{2010arXiv1005.1831B}. These models must be considered carefully as they may change the predictions for the positron fraction within the candidate PWN.

Secondly, and more importantly, the termination shock that accelerates e$^+$e$^-$ pairs to high energies would also trap the accelerated e$^+$e$^-$ within the PWN~\citep{1984ApJ...283..694K, 2001A&A...380..309V}. Due to the high energy density of the magnetic and radiation fields within the PWN, these leptons may quickly lose energy if they are not able to escape into the interstellar medium. To avoid this constraint, \citet{2009JCAP...01..025H} indicated``mature" (middle-aged) pulsars as the source of the rising positron fraction. These systems include a pulsar which is currently moving through the termination shock, forming a bow-shock nebula at which additional particle acceleration can take place. This particle acceleration is unlikely to be confined within the PSR due to the pulsar motion, providing an outlet for  for high energy leptons to escape from the confining magnetic fields of the PWN. Additionally, recent observations by \citet{2011ApJ...743L...7H} indicate that the lack of TeV $\gamma$-rays observed from the Vela X PWN may be due to the effective escape of these particles out of the PWN, providing the first evidence that high energy electrons may efficiently leave the PWN before losing significant energy. One concern for this scenario is that the majority of the e$^+$e$^-$ flux injected by the natal pulsar be be radiated away before the bow shock allows for leptons to escape. 

Finally, an analysis by \citet{2009PhRvD..80f3005M} pointed out the alternative possibility that significant e$^+$e$^-$ acceleration occurs at the shock-front of the SNR with the ISM. This can greatly decrease the contribution of pulsars to the observed positron fraction, since magnetic fields in this region are likely to primarily accelerate electrons. 

\section{Polarization Measurements of PWN}
\label{sec:polarization}

These uncertainties indicate that a new method is necessary in order to understand the lepton population inside PWN. One important characteristic of PWN is the high linear polarization fraction observed across PWN sources, often contributing a fractional intensity of 30-50\%~\citep{2012SSRv..166..231R}. Since the maximum linear polarization for a power-law electron spectrum producing a synchrotron spectral index $\alpha$ is (3$\alpha$+3)/(3$\alpha$+7), the relatively flat synchrotron spectral indices of 0.0--0.6 imply that the energy density of the ordered magnetic field in the PWN often significantly exceeds half of the total magnetic field energy density. The high magnetic field order in these regions implies that their synchrotron emission should additionally have observable circular polarization. 

In this work, we follow the derivation of \citet[][hereafter WW97]{1997ApJ...475..661W} and calculate the circular polarization of synchrotron radiation stemming from a power-law distribution of negatively charged electrons to be \citep{1968ApJ...154..499L}:

\begin{equation}
\label{eq:circular}
\left(\frac{V}{I}\right) = \frac{4}{\sqrt{3}} \frac{b(\gamma)}{a(\gamma)}\cot(\theta)\sqrt{\frac{qB_0 \sin(\theta)}{2\pi m_e c f}}
\end{equation}

where V is the intensity of circularly polarized synchrotron radiation, I is the total synchrotron intensity, $\gamma$ is the index of the power-law spectrum for the relativistic leptons, $\theta$ is the angle between the direction of an ordered magnetic field and the line of sight (assumed to be $\pi$/4 in all that follows), q is the charge of an electron, B$_0$ is the magnetic field strength, m$_e$ is the mass of an electron, c is the speed of light, $f$ is the observed synchrotron frequency and b($\gamma$) and a($\gamma$) are parameters calculated by \citet{1968ApJ...154..499L}. 

The decrease in the observed circular polarization due to magnetic field disorder can be modeled as a decrease in the Stokes V parameter  V$_{obs}$~=~d$_c$V$_{emit}$, where V$_{emit}$ is the intrinsic circular polarization given by Equation~\ref{eq:circular}, V$_{obs}$ is the observed circular polarization, and 0~$<$~d$_c$~$<$~1 characterizes the depolarization due to the magnetic field. Notably, this scheme can also be employed for the linear polarization with a depolarization parameter we mark as d$_l$, and models by \citet{1979MNRAS.186..519B} find that for standard structures of magnetic field turbulence, d$_c$~$\sim$~g~d$_l$ with g$~\approx~$0.9. 

Thus, as noted by WW97, the observed linear polarization in PWN can be used to approximate the expected circular polarization within the PWN. For multiple PWN, observations of linear polarization on the order of 30-50\% can be translated to coefficients of d$_c$~$>$~0.5. We note that d$_c$ may be significantly lower in specific PWN, most notably the Crab, where the observed linear polarization is approximately 8.5\% \citep{1969ApJ...158..145J,1966ApJ...144..437B}, yielding a measurement of d$_l$ of 0.17-0.25 (WW97). However, in any specific PWN, the degree of linear polarization is measurable, allowing a direct calculation of the expected circular polarization.

There are three caveats to this argument, as again first pointed out by WW97. First, the magnetic field could be preferentially aligned perpendicular to the direction of observation, making cot($\theta$) small and the circular polarization unobservable. However, this can be tested through observations of several PWN filaments, or even multiple observations across a single PWN. Second, the magnetic field turbulence could be produced in such a way as to destroy the circular but not the linear polarization. Third, and most interestingly, the synchrotron radiation could be produced by equal populations of positrons and electrons, the opposite charges of which cancel their contributions to the circular polarization exactly, while contributing constructively to the linear polarization.

\begin{figure}
                \plotone{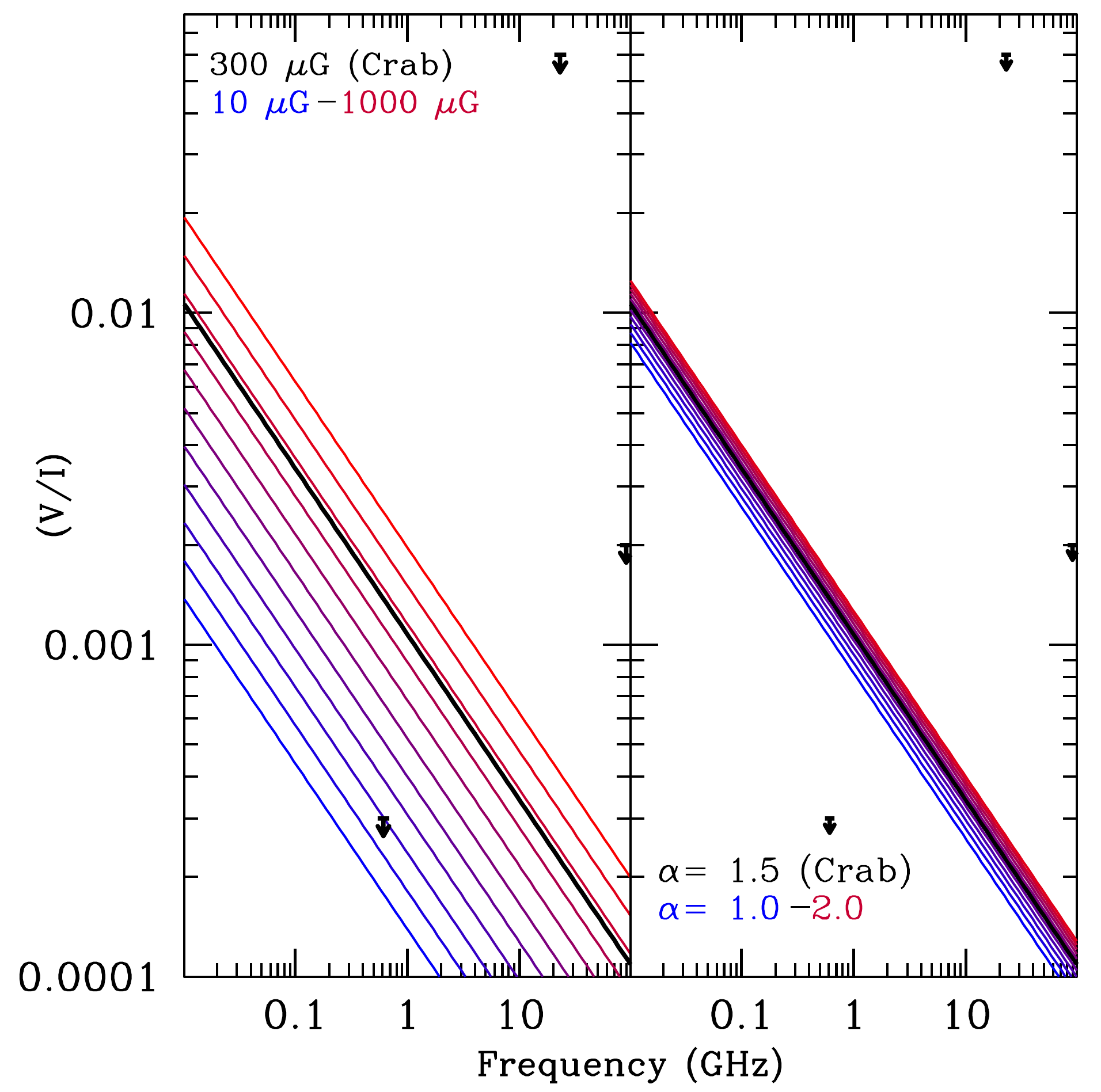}
                	\caption{ \label{fig:polarization} Circular polarization as a function of the observed radio frequency for (Left) a pure electron population with a power-law spectrum $\frac{dN}{dE}$~=~-1.5 in an ordered magnetic field of strength 10 $\mu$G (blue) and 1000 $\mu$G (red), the equipartition value for the Crab nebula of 300 $\mu$G is shown in bold black, and (Right) for a power-law spectrum with index $\frac{dN}{dE}$~$\propto$~E$^{-\alpha}$ with $\alpha$ linearly distributed between 1.0 (blue) and 2.0 (red), the assumed Crab value of $\alpha$~=~1.5 is shown in bold black. In both panels limits for the Crab nebula are shown at 610~MHz~\citep{1997ApJ...475..661W}, 23~GHz~\citep{1980ApJ...239..873W}, and 89.2~GHz \citep{2011A&A...528A..11W}. Note that the models for V/I are shown assuming a perfectly ordered magnetic field (d$_c$~=~1.0), and must be decreased by the value of d$_c$ in order to compare with the observations. We find that the circular polarization anticipated in multiple PWN can be detectable with current radio instruments so long as either the magnetic field strength is high, or the magnetic field is relatively ordered (i.e. d$_c$~$\approx$~1).}
\end{figure}

\begin{figure*}
                \plottwo{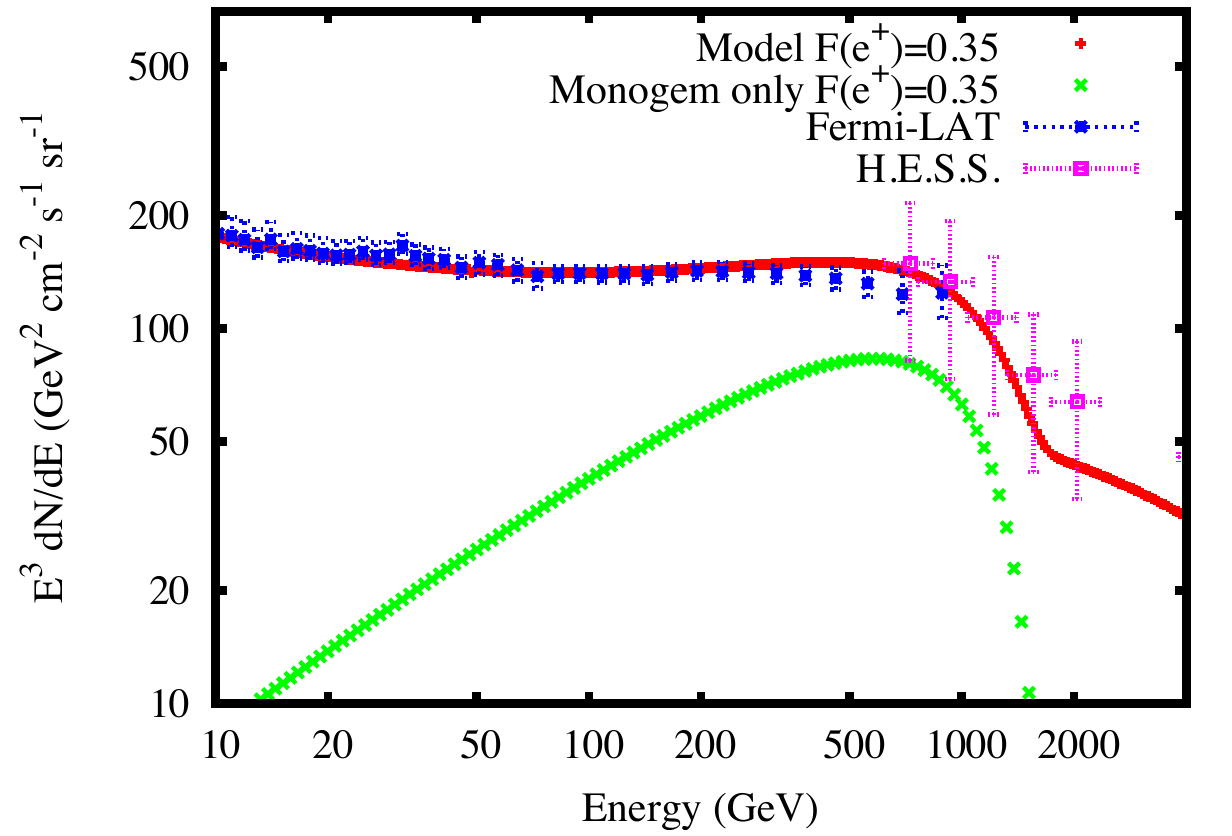}{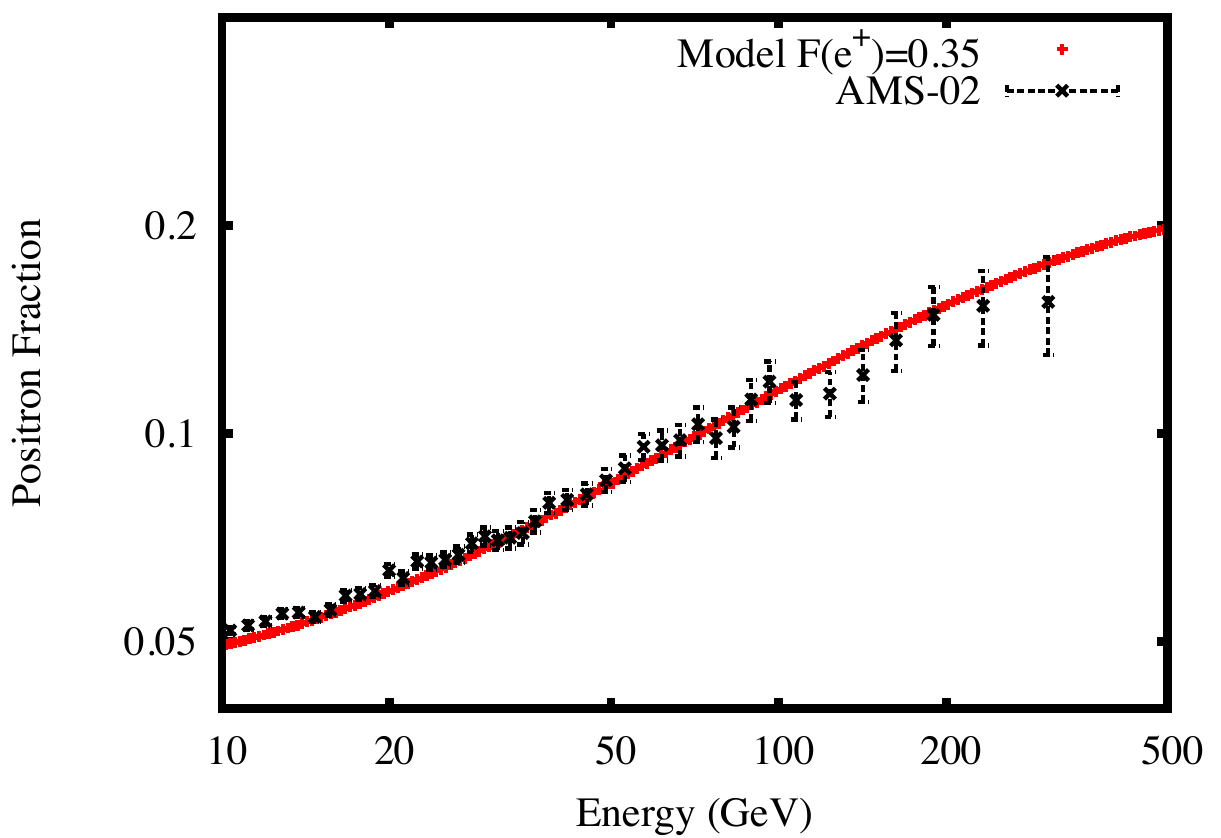}
                	\caption{ \label{fig:positron_fraction} Models for the total lepton spectrum and positron fraction from models where the positron fraction is dominated by the nearby Monogem PWN, which is producing e$^+$e$^-$ emission with a positron fraction, F(e$^+$) of only 0.35, compared to observations by Fermi-LAT, H.E.S.S. and AMS-02. All other values for the modeled PWN emission and the Galprop astrophysical background~\citep{1998ApJ...509..212S} follow those from \citet{2013ApJ...772...18L} except the astrophysical electron injection spectrum is taken to be 2.55, rather than 2.5, and the total pulsar energy deposited into e$^+$e$^-$ is taken to be 1.3~$\times$~10$^{49}$~erg rather than 8.6~$\times$~10$^{48}$~erg.}
\end{figure*}

Several measurements have been made for the circular polarization of the Crab PWN, and while no circular polarization has yet been detected, upper limits on the fractional circular polarization have been set at 0.03\% at 610 MHz~\citep{1997ApJ...475..661W}, 6\% at 23~GHz~\citep{1980ApJ...239..873W}, and 0.2\% at 89.2~GHz~\citep{2011A&A...528A..11W}. In Figure~\ref{fig:polarization} we plot these limits alongside the predicted circular polarization fraction for the Crab Nebula for varying values of the magnetic field strength within the PWN varying from 10-1000~$\mu$G, and assuming no depolarization from the magnetic field structure (i.e. d$_c$~=~1.0). The equipartition magnetic field of the Crab nebula is estimated to be 300~$\mu$G \citep{1984ApJ...278L..29M}. However, while the circular polarization at a given frequency scales as $\sqrt{B}$, the circular polarization as a function of frequency scales as 1/$\sqrt{B}$. This means  that circular polarization measurements can be constraining for any magnetic field model, so long as a sufficiently low observation frequency is picked. We note in Figure~\ref{fig:polarization} (right) that the assumed spectral index has only a negligible effect on the predicted circular polarization.

Using these constraints, WW97 argued that the Crab PWN likely contains a non-negligible positron population. However, from observations of only the Crab nebula, they are unable to entirely rule out the possibility that the magnetic fields of the Crab are aligned such that the expected circular polarization is extremely small. In what follows, we note that this argument can be significantly strengthened with current instruments and observations. 

\section{Low-Frequency Observations}

Observations by the Low-Frequency Array For Radio Astronomy \citep[LOFAR,][]{2013A&A...556A...2V} telescope may significantly enhance our ability to observe circular polarization from PWN. The LOFAR telescope provides unprecedented sensitivity and angular resolution for very low frequency radio observations in the range spanning 15 -- 240~MHz, and is capable of processing the full Stokes IQUV intensities. For the observation at hand, LOFARs low-frequency observational bands significantly strengthen the expected circular polarization signal, which falls as 1/$\sqrt{f}$. Additionally, low-frequency observations are critical for the detection of circular polarization, due to the possibility of contamination from relativistic Faraday rotation of an initially linearly polarized synchrotron signal, an effect which varies linearly with the observational frequency~\citep{1998PASA...15..211K}. Returning to Figure~\ref{fig:polarization}, we note that at frequencies near 60~MHz, the expected circular polarization of the Crab Nebula is modeled to be approximately 0.44\%, a factor of 3.2 higher than at 610~MHz. This allows for stronger constraints to be set on the circular polarization of the Crab Nebula, which will strengthen the constraints on the positron fraction inside the PWN.

While circular polarization measurements are extremely difficult due to systematic issues and atmospheric effects, we note that if 60~MHz LOFAR observations are ultimately capable of placing a limit on the circular polarization in the Crab nebula approximately as strong as the 610~MHz observations of WW97 (V/I~$<$~0.03\%), the positron fraction inside the PWN would be observationally constrained to be at least 0.36.  

While traditional models of the positron induction by pulsars assume (for strong theoretical reasons) an injected positron fraction of 0.5, in principal there is no reason why the injection of primarily leptons with a smaller positron fraction would not be capable of producing AMS-02 observations. In Figure~\ref{fig:positron_fraction}, we produce a model where the Monogem pulsar injects primary electron/positron pairs with a positron fraction of only 0.35, and compare these results to the total e$^+$e$^-$ spectrum observed by the Fermi-LAT~\citep{2010PhRvD..82i2004A} and H.E.S.S.~\citep{2008PhRvL.101z1104A} telescopes, and the rising positron fraction observed by AMS-02~\citep{2013PhRvL.110n1102A}. We find that these produce a reasonable fit to both observations, indicating that circular polarization measurements can produce a sufficient constraint on the positron fraction within PWN to support the PWN interpretation of the rising positron fraction.

\section{Observations of Bow-Shock Nebulae}

Additionally, the high sensitivity of LOFAR allows for the possible observation of circular polarization in multiple other PWN, which may prove to be superior targets for circular polarization studies. In order to constrain models of the rising positron fraction, observations of bow-shock PWN are especially critical, due to the fact that bow-shocks are thought to be necessary in order to allow positrons produced in PWN to leak into the interstellar medium~\citep{2011heep.conf..624B}.

One compelling target is G189.22+2.90, a bow-shock PWN associated with the SNR IC 443 and with an age of 30~kyr~\citep{1999ApJ...511..798C, 2001ApJ...554L.205O}. In this system, the pulsar is currently propagating through the reverse shock, which has a characteristic magnetic field of approximately 20$\mu$G \citep{1997ApJ...490..619S}. While this low magnetic field decreases the expected circular polarization in the LOFAR frequency range, this is offset by the fact that the observed linear polarization in G189.22+2.90 is relatively large, with a lower limit of 8\%, and with several observed regions exceeding 25\% \citep{2001ApJ...554L.205O}. This indicates a value of d$_l$~$>$~0.25, and likely as high as 0.5. In the case that d$_c$~=~0.5, LOFAR observations at 60~MHz with a sensitivity to circular polarization of 0.02\% would be able to constrain the e$^+$e$^-$ fraction to exceed 33\% in the PWN.  

\begin{figure}
                \plotone{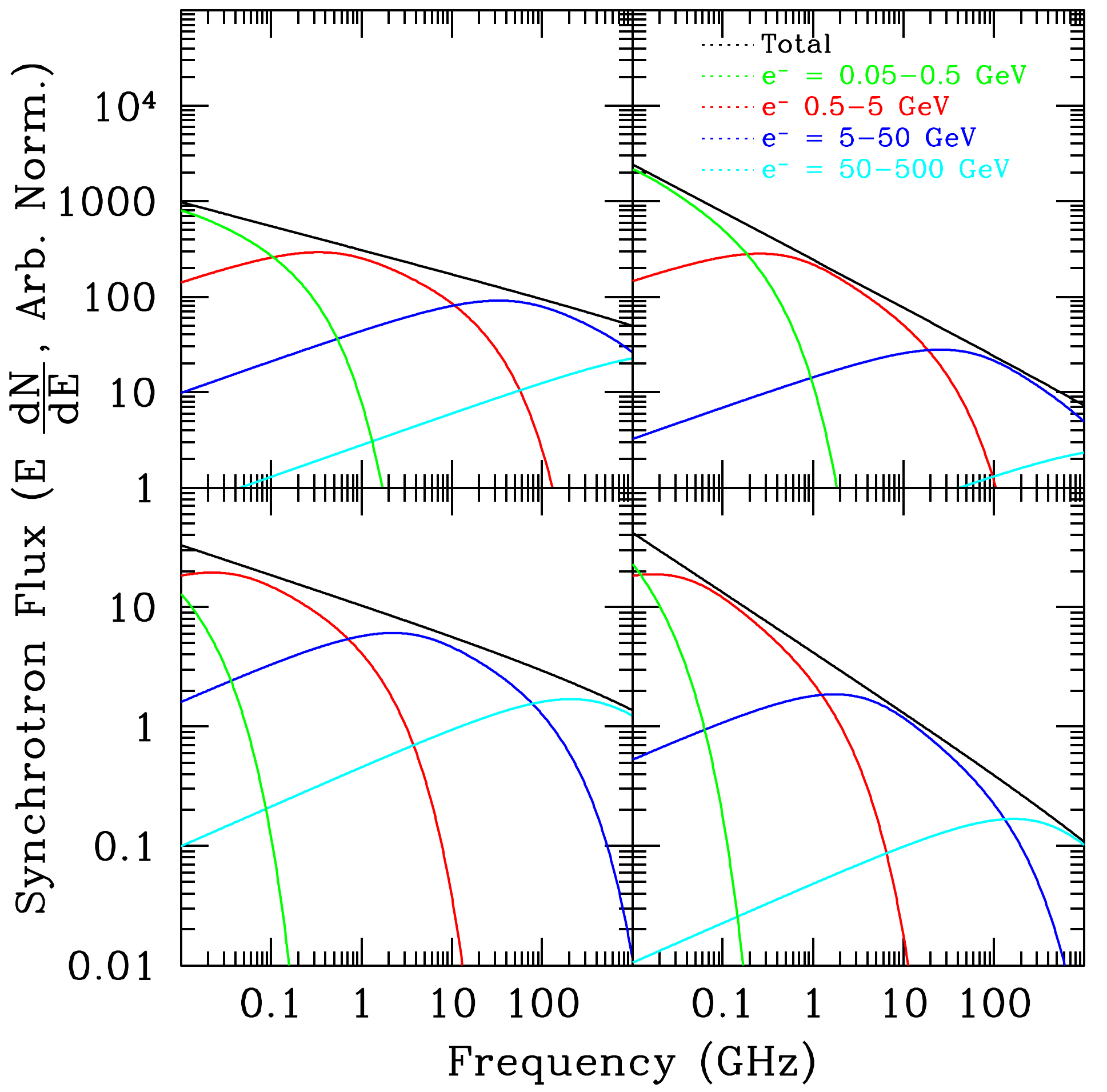}
                	\caption{ \label{fig:electronenergies}  Total Synchrotron Emission (black solid) as a function of frequency for models with a magnetic field strength of 300$\mu$G (top) and 10$\mu$G (bottom) for two different choices of a power-law electron injection spectrum: $\frac{dN}{dE}$~$\propto$~E$^{-1.5}$ (left) and $\frac{dN}{dE}$~$\propto$~E$^{-2.0}$ (right), along with the fractional contribution from electrons in the energy range 0.05--0.5 GeV (green) 0.5-5 GeV (red), 5-50 GeV (blue), and 50-500 GeV (cyan).}
\end{figure}

Furthermore, the low-magnetic fields observed in G189.22+2.90 indicate that the electron population dominating the synchrotron radiation in the LOFAR frequency range is of significantly higher energy than for the Crab PWN. In Figure~\ref{fig:electronenergies} we show the total synchrotron emission spectrum for models with magnetic fields of 300~$\mu$G and 20$\mu$G, breaking down the contributions among electrons of different energy ranges. We find that in the case a 20$\mu$G magnetic field, the emission at 60~MHz is dominated by electrons in the energy range of 0.5--5~GeV, rather than by electrons in the range of 50--500~MeV, as in the case of the Crab PWN. This stands as an important test of the pulsar positron hypothesis, since circular polarization observations in low magnetic field regions brights us closer to the 10-300~GeV energy range characteristic of the observed AMS-02 excess. Moreover, thermally produced electrons may contribute significantly to the low energy lepton flux, creating an e$^+$e$^-$ asymmetry, which would cutoff at lower energies than the observed AMS-02 excess~\citep{2011heep.conf..624B}.

\section{Discussion and Conclusions}
\label{sec:conclusions}

In this work, we have shown that a null observation of circular polarization in the Crab PWN (and potentially other luminous nebulas) would provide a strong indication that the positron fraction within PWNs is significant. This not only has important implications for our understanding of the PWN environment, but also implies that positrons from PWN exist in the quantity necessary to produce the rising positron fraction observed by PAMELA and AMS-02. Notably, telescopes like LOFAR have the capability to produce high sensitivity circular polarization measurements at extremely low frequencies $\sim$100~MHz, where the circular polarization signal is maximized.  In observations of the Crab Nebula, AMS-02 observations are capable of showing that the positron fraction in the PWN is at least 0.36.

Additionally, this method has the capability to go beyond ruling out pulsar interpretations of the AMS-02 excess. Radio surveys of more mature, ``bow-shock" PWN show an elongated radio tail, which does not appear in X-Ray observations (e.g. the Mouse PWN \citep{2004ApJ...616..383G} and G189.22+2.90). In these systems, the e$^+$e$^-$ observed near the termination shock are expected to freely diffuse into the interstellar medium and produce a portion of the interstellar lepton flux. While the magnetic fields in these mature pulsars are smaller than within the Crab PWN, making the strength of the circular polarization signal smaller, this effect also increases the typical e$^+$e$^-$ which dominates synchrotron production within the LOFAR radio band, with the final effect that the electrons producing the synchrotron emission are of a similar energy to those implicated in AMS-02 rising positron fraction. Future studies may be able to map the local positron fraction in regions of higher synchrotron intensity, constraining the escape of e$^+$e$^-$ from PWN and providing a strong indication that pulsars are responsible for the rising positron fraction.

\acknowledgments
TL thanks Michiel Brentjens, Pasquale Blasi, Ilias Cholis, Ger de Bruyn, Dan Hooper, Emmanuela Orru' and Farhad Yusef-Zadeh for helpful discussions. TL is supported by the National Aeronautics and Space Administration through Einstein Postdoctoral Fellowship Award Number PF3-140110. This work was supported in part by the National Science Foundation under Grant No. PHYS-1066293 and the hospitality of the Aspen Center for Physics. \\

\bibliography{polarization} 

\end{document}